\documentclass[sn-mathphys]{sn-jnl}

\begin{document}

\title[Article Title]{Progress on levitating a sphere in cryogenic fluids}

\author[1]{\fnm{M.} \sur{Array\'as}}
\author[2]{\fnm{F.} \sur{Bettsworth}}
\author[2]{\fnm{R.~P.} \sur{Haley}}
\author[2]{\fnm{R.} \sur{Schanen}}
\author[1]{\fnm{J.~L.} \sur{Trueba}}
\author[1]{\fnm{C.} \sur{Uriarte}}
\author[2]{\fnm{V.~V.} \sur{Zavjalov}}
\author[2]{\fnm{D.~E.} \sur{Zmeev}}

\affil[1]{\orgdiv{Área de Electromagnetismo}, \orgname{Universidad Rey Juan Carlos}, \orgaddress{\street{Tulipán s/n}, \city{Mostoles}, \postcode{28933}, \state{Madrid}, \country{Spain}}}

\affil[2]{\orgdiv{Department of Physics}, \orgname{Lancaster University}, \orgaddress{\street{Physics Building}, \city{Lancaster}, \postcode{LA1 4YB}, \state{Lancaster}, \country{United Kingdom}}}


\abstract{We present the working prototype of a levitation system designed for investigation of flows in cryogenic helium fluids. The current device allows the levitation of a superconducting sphere and has several provisions made for allowing precise control over its motion. We report on progress in the detection and control systems of the prototype and demonstrate how uniform circular motion can be implemented.}




\maketitle

\section{Introduction}\label{sec1}
A superfluid flow is influenced by quantum effects and, as is the case of the classical turbulence, quantum turbulence is not fully understood. A basic problem of practical importance is the  determination of the high-Reynolds number flow produced by a sphere moving steadily through a fluid at rest. Turbulence develops in the vicinity of the solid body and evolves at large distances from the sphere \cite{Batchelor}. Viscous interaction with the surface of the sphere perturbs the flow and leads to the formation of boundary layers. The boundary layer becomes unstable behind the sphere, the flow undergoes a separation and a turbulent wake develops. A fundamental problem is how this picture must be modified for a superfluid \cite{Skrbek21}. Currently, there is no definite picture of the structure of the boundary layer and the boundary condition at the solid-superfluid interface in the limit of low temperatures. However, there is circumstantial evidence that the condition is of slip type \cite{Zmeev2015, Zieve2012}. 
The study of quantum turbulence poses challenging difficulties, both experimentally and theoretically. There has been almost no experimental studies of boundary layers and separation in quantum turbulence, and the visualization techniques available for classical fluids are not directly applicable to the quantum case \cite{Vinen02}. This is just one example of a problem that careful measurements on a body of regular geometric shape moving in a superfluid can resolve.

Here we report the progress on the implementation of a novel device designed to explore the motion of a sphere in a superfluid \cite{Arrayas21}. The device was conceived for experiments on both superfluid $^4$He and $^3$He, but direct comparisons with cryogenic fluids above the critical temperature, $T_c$, would be instrumental in understanding the differences between classical and quantum turbulence. For experiments in $^4$He the goal is to extend the celebrated experiments of Wilfried Schoepe and his group on a levitated sphere above a superconducting plane \cite{Schoepe95, Schoepe22}. It promises to open a new chapter in the investigation of moving objects within stationary superfluid because, unlike earlier work, the researcher will have total control of the movement of the sphere within a 2D plane. Thus we can investigate, not only to-and-fro oscillatory motion as previously \cite{Vinen14} but in addition, for example, uniform rectilinear motion and steady motion in a circle.

The uniform rectilinear motion has been discussed previously \cite{Arrayas21}. Here we introduce the detection and control systems and explain how they can be used to induce uniform circular motion for the sphere. We also demonstrate that levitation within the design outlined in our earlier work \cite{Arrayas21} is indeed achievable.
\section{Levitation system}
Here we summarize the main components and some technical details of our working prototype. The prototype is built largely based on the design proposed in \cite{Arrayas21}.

Our device is made up of 6 independent coils. Two of these coils (with radii $r_{in}=6$\,mm; $r_{out}=15$\,mm) have a concentric configuration with their axes coinciding and both are located on a horizontal plane. These two coils are responsible for producing the levitation of the In sphere and will produce the most intense field of the system. The number of turns in each of the coils is 250 and the currents through each must be carefully adjusted to a ratio of $I_{in}/I_{out}=1.2$. In addition, both coils are supplied with currents in opposite directions in such a way that in the plane $z>0$ (which is the region delimited by the 6 coils, where it is intended to levitate the sphere), the vertical component of the field created by the internal coil would be positive, while the vertical component of the field of the outer coil would be negative.

The other four coils (let us call them lateral coils, each with radius $r_{lat}=6$\,mm) are responsible for controlling the motion of the sphere. Through them it is possible to drive the sphere in the different movement regimes inside the superfluid. These coils are located in vertical planes (perpendicular to the plane defined by the levitation coils) and are also placed in pairs of coils, two of them with their axis parallel to one of the directions of the horizontal plane (the $x$ axis) and the other pair of coils with their axis perpendicular to the first pair (the  $y$ axis). The centres of the four coils are located in a horizontal plane that is elevated with respect to the plane of the levitation coils by a height $z=5$\,mm. Each pair of coils of the same axis are separated between their centres 2\,cm, in such a way that the projection of their planes with the plane of the levitation coils constitutes a straight line contained in the space delimited by the concentric levitation coils. Each of these four lateral coils is wound 250 turns and the currents through them will depend on the regime of movement that is intended to be obtained in the sphere.

The levitating sphere is formed by a 1\,mm diameter polyacetal (Delrin) core on which a thin layer of In 40\,$\mu$m thick is coated. This gives the sphere an average density of 2.33\,g\,cm\,$^{-3}$ with a mass of 1.7\,mg. Due to this, to establish the equilibrium of the sphere at a height of 5mm (this height is defined by the position of the lateral control coils) above the levitation coils, it is necessary to set the currents of the levitation coils at 2.64\,A (inner levitation coil) and 2.2\,A (outer levitation coil).

The above description refers to an ideal design obtained through various numerical and analytical calculation models. In these models, we used several simplifications that are not feasible from an experimental point of view, so it is necessary to adapt various features of the model prior to its construction.

On one hand, for the calculation of the magnetic fields, it was assumed that the coils were ideal, and the successive turns of the coils do not occupy much height. However for the manufacture of the prototype, it is needed to wind each of the coils minimizing the section occupied by the wires. For this reason, a maximum section of wire in each coil of 4\,mm$^2$ was assumed, winding the 250 turns of each coil inside a 2$\times$2\,mm$^2$ toroidal groove. We used NbTi alloy superconducting wire with a cross-section of 105\,$\mu$m for winding all coils. This wire allows to wind the 250 turns through 15 layers with 15 to17 turns in each one. In addition, each layer of wires must have an almost perfectly tight distribution, which implies the necessity to wind all the coils manually, to ensure maximum homogeneity of the magnetic field. For this purpose we used a winding machine designed and built in-house.

\begin{figure}
  \centering
  \includegraphics[width=0.25\textwidth]{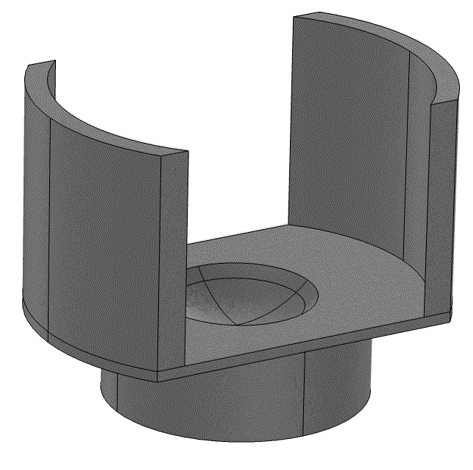}\\
  \includegraphics[width=0.35\textwidth]{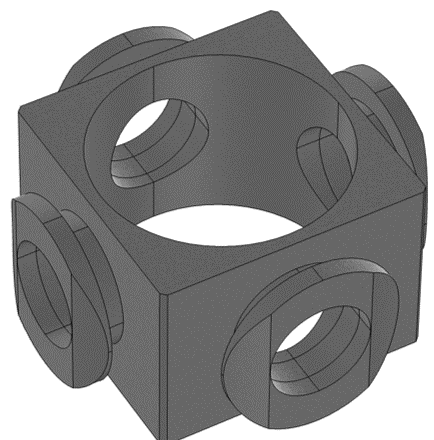}\\
  \includegraphics[width=0.44\textwidth]{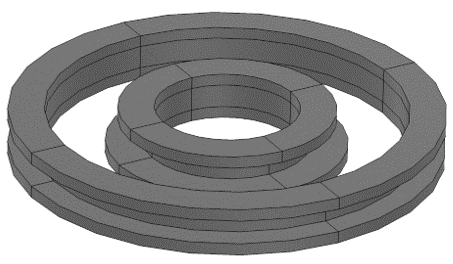}
  \caption{The exploded view of the GRP frame. From top to bottom: launching platform, lateral coils holder, levitation coils formers. }\label{frame}
\end{figure}

All these coils were wound on a GRP (glass reinforced polymer) frame. The frame is divided into different parts. The main part of the device is the lateral coils holder as depicted in Fig.~\ref{frame}. This part consists in a hexahedron with 4 cylindrical attachements on its lateral faces. These attachements have  2$\times$2\,mm$^2$ machined toroidal grooves (with a square cross-section) on which the side coils are wound and have an inner cylindrical hole of 4\,mm radius to provide optical access to the sphere inside the levitation chamber, which is defined by the space limited by the hexahedron. On the upper and lower faces, the lateral coils holder has a larger cylindrical cavity of 8\,mm radius through which the levitation coils (in the lower cavity) and the sphere levitation platform (upper cavity) are inserted.

The levitation coils are wound on two additional separated formers shown in Fig.~\ref{frame}. These formers consist of cylindrical parts of GRP with toroidal grooves of square cross-section 2$\times$2\,mm$^2$ machined on their lateral surfaces. The inner coil former is inserted into the lower cavity of the hexahedron, and is perfectly flush due to an existing flap on the lower base of the inner coil frame. The interior cavity of the frame of the external coil is adjusted to the lateral coils formers. 

\begin{figure}
  \centering
  \includegraphics[width=0.6\textwidth]{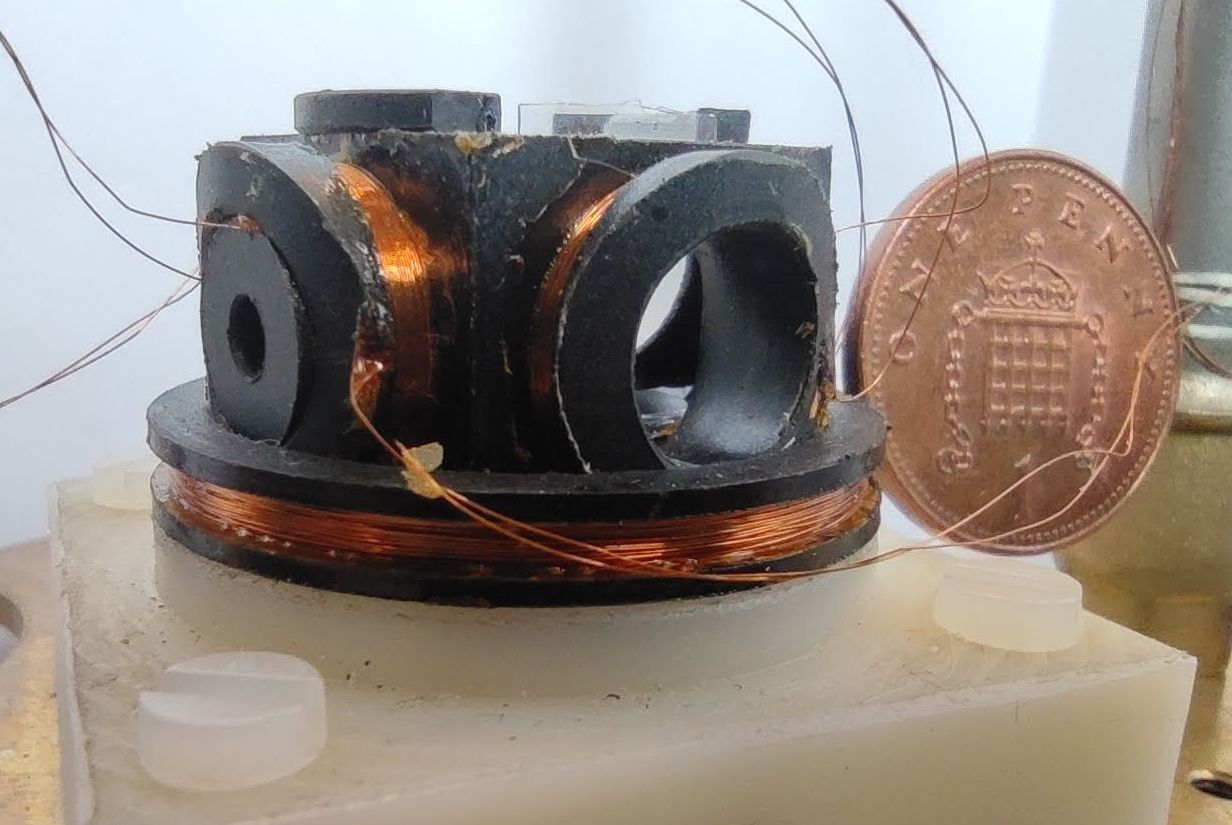}
  \caption{The first working prototype of the levitation system. A penny coin is shown for scale.}\label{realone}
\end{figure}

The launching platform which can be seen in Fig.~\ref{frame} is inserted in the upper cavity of the hexahedron. The platform consists of a goblet-shaped piece intended to place the sphere initially close to the expected equilibrium position ($x=0$, $y=0$, $z=5$\,mm) before levitation. This platform has a solid cylindrical base that inserts directly into the cavity of the inner levitation coil former. The region where the sphere is placed has concave geometry to prevent possible initial displacement of the sphere. Two flat glass sheets are glued to its sides to constrict the movement of the sphere to the interior of the hexahedron and at the same time provide optical access. Finally, the actual assembled device is shown in Fig.~\ref{realone}.

\section{Fabrication of  indium sphere}

As we explain above, a solid ball made from uniform superconductor would be too heavy to levitate. The requirement of low average density is somewhat relaxed if buoyancy is taken into account. We can satisfy the low-density requirement by using a plastic bearings ball plated with a thin layer of superconductor. The field distribution necessary for the levitation holds as long as the plated layer is much thicker than the London penetration depth (typically, 100\,nm). We chose 1\,mm diameter Delrin ball bearings with the density of 1.35\,g\,cm$^{-3}$. In order to make the balls conductive, we first plate them with a 100\,nm thick layer of silver using Tollen test reaction, also known as the silver mirror reaction \cite{Chem_textbook}. After this, the balls can be electroplated with a desired superconductor element. We chose indium as type I superconductor to avoid flux trapping, which would hinder reproducibility of the motion.

	Electrodeposition of micro-scale indium layers is not new to low temperature applications, it has been used in the fabrication of on-chip refrigeration \cite{Sarsby-500microkelvin-2020, Yurttagul-indiumasa-2019} and in the making superconducting heat switches \cite{Bhattacharyya-indiumswitch-1982}. Deposition for our intended purpose however requires solving the non-trivial challenge of uniformly coating a sphere. Electroplating deposits material on any surface charged oppositely to the aqueous metal ions and our spheres must be in electrical contact with a cathode to maintain this charge. If the spheres were left to rest on a flat metallic cathode, metal would be primarily deposited on the hemispherical surface furthest from, along with depositing on, the cathode itself. Minimal if any deposition should occur on the closer surface and no layer will grow where the spheres and cathode meet. There is also a risk that indium grown on the sphere will join with that on the cathode, fusing them together. This imbalance and sticking could be alleviated through stirring of the solution during deposition but even light stirring will lift the spheres off the cathode, removing a necessary electrical connection.
	
	Our silver-coated spheres were laid to rest on a strip of silver foil which acted as the cathode in this process. To even out the indium deposit we periodically jostled the foil to roll the spheres and expose the less plated areas. The deposition was performed with a bath chemistry of 0.26~mole In(SO$_3$NH$_2$) in aqueous sulfamic acid. We chose pulsed direct voltammetry to allow for metal ion concentration to replenish between growth cycles and for more uniform deposition \cite{Tian-electrodepositionindiumbumps-2010}. The samples were plated with a forward current of 0.1\,mA for 2\,ms with a target density of 20\,mA\,cm$^{-2}$ and then left to rest with zero current for 8\,ms. The total plating time was 900\,s which resulted in an indium layer of thickness 40\,$\mu$m, estimated from mass measurements.
	
	In our preliminary experiments we did not pay an attention to the quality of the surface. However, the surface quality is crucial for quantitative experiments. We believe that by carefully choosing the parameters of electroplating or by polishing we can achieve surfaces with roughness much smaller than what has been common in the experiments with  objects moving in cryogenic fluids so far.

\section{Detection system}
In order to make experiments in $^3$He and $^4$He at ultralow temperatures, we need to design a reliable detection system for the movement of the sphere inside opaque cryostats. 
The proposed solution is based on the property of the perfect diamagnetism behaviour of superconductors. If a perfect diamagnetic object moves through a magnetic field, it expels the field lines from its interior. A detection coil can be used to measure the flux variations due to the relative position of the sphere with respect to the coil. These flux variations can be measured through the impedance of the detection coil when fed by a fixed high-frequency AC current. The impedance is proportional to the frequency $\omega$ and to the inductance of the detection coil $L$. The inductance of the coil depends on the position of the sphere, so we can correlate the measured impedance with the position of the sphere in the experimental cell and verify the measurement in preliminary experiments in an optical refrigerator. 

To test this proposal, we created a model to calculate the sensitivity and parameters involved in the building of the detection coils, identifying those that maximize the voltage induced on the detection coil.  The model consists  of a superconducting sphere and two concentric coils. One of the coils is responsible for creating the quasistatic magnetic field, and is supplied with a static current of 2\,A (similar to the coils present in our experiment) and has 250 turns. Concentric to this, the detection coil is located in the same plane and is supplied with an AC current of amplitude $0.01$\,A at a frequency of 10\,MHz. Those values are chosen in order to maximize the induced voltage without interfering with the movement of the sphere.

\begin{figure}
  \centering
  \includegraphics[width=0.4\textwidth]{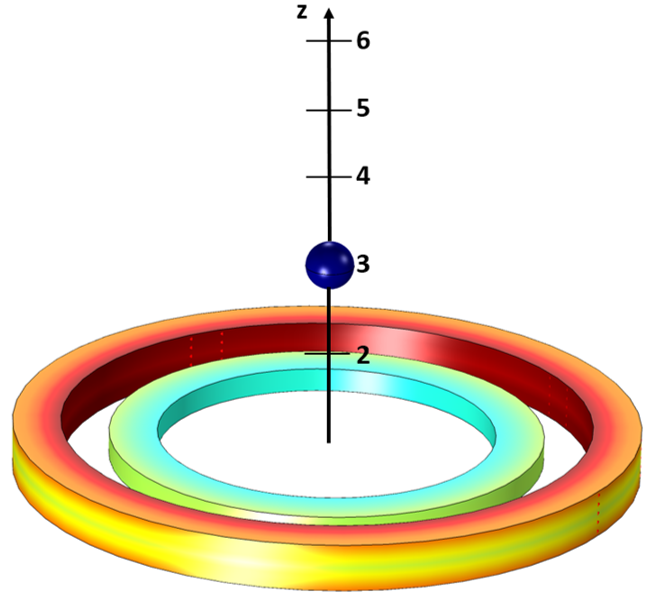} 
  \includegraphics[width=0.5\textwidth]{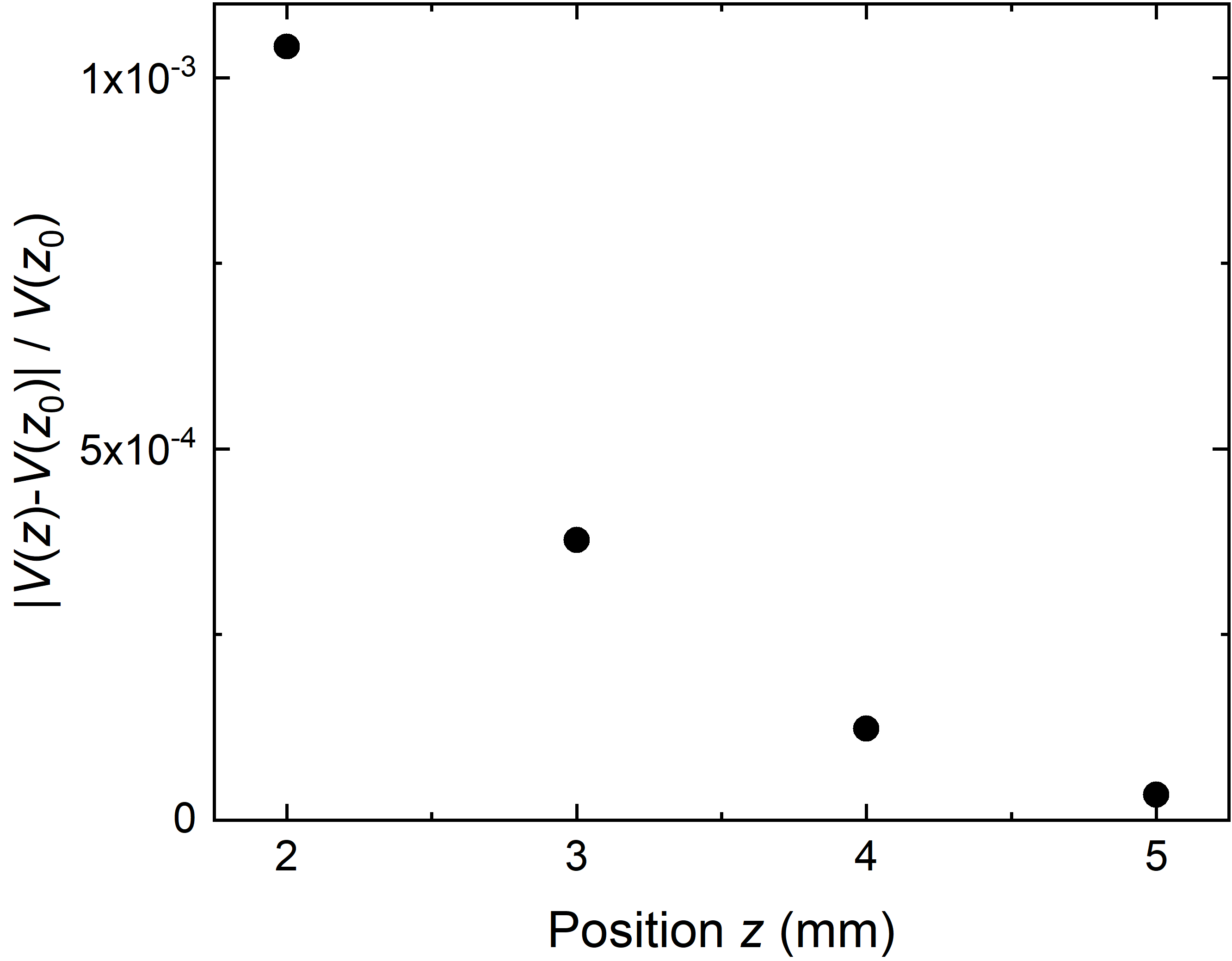}
  \caption{\emph{ Left:} A schematic representation of the model for testing the sensitivity of the detection coil (inner coil) to the position of the sphere along $z$. \emph{Right:} Absolute value of the relative change of the voltage $V(z)$ across the detection coil at different positions $z$ with respect to the position $z_0$=6\,mm.}\label{detection}
\end{figure}

The radius of the outer coil (the one supplied with DC current) is 6\,mm and the radius of the detection coil is set to 3\,mm. Our superconducting sphere is located at a point on the $z$ axis of the coils, separated by a certain distance. A schematic representation is shown in the left-hand side of Fig.~\ref{detection}.

We have carried out several studies to determine the detection sensitivity of the system. Specifically, it is of special interest to verify if it is possible to distinguish between two different positions of the sphere when it is stationary. For this purpose, we have calculated using finite elements analysis the magnetic flux through the detection coil so we can calculate the voltage across the detection coil when the sphere is at a particular position. We take $z = 6$\,mm as a reference position.  The result is shown in the right-hand side of  Fig.~\ref{detection}. Bigger sensitivities can be achieved with the number of turns of the coil and increasing the current through it. In addition, the ratio of the radius of the coil to the radius of the sphere must be as small as possible and the cross-sectional area of the detection coil should be minimized. This requirement needs to be reconciled with the requirement for unobstructed motion of the sphere far enough from the walls.

\section{Control of the sphere in circular uniform motion}
While oscillatory motion of the probes is perhaps the most studied so far in quantum turbulence experiments and theories, our design allows for other modes of motion as well. In previous work \cite{Arrayas21}, we have suggested the method for driving the sphere in uniform rectilinear motion, based on \cite{Zmeev14}. Here we present the control method to drive the sphere in circular uniform motion about the vertical axis and some simulations performed to test it.

Both pairs of lateral coils are used here.  The idea is to let the sphere initially reach the equilibrium without introducing any control currents (other than levitation currents). Subsequently, using one of the pairs of control coils, an identical narrow pulse is introduced in both coils of the pair, in such a way that the sphere leaves its equilibrium position up to a given amplitude (this amplitude will be a function of the dimensions of the orbit that is intended to be obtained, as well as the speed of the sphere desired in circular motion) and moves along one of the axis (let say the x axis). The equation for the pulse takes the form,

\begin{equation}
	\label{eq:movcircular}
	I_{L,x}(t) = \left\{ \begin{array}{lcc}
		1.45\sin(180\pi(t-t_0)) & \mbox{if} & t_0 < t \leq t_0+0.005, \\
		\\ 0 & \mbox{otherwise},\\
	\end{array}
	\right.
\end{equation}
where $I_{L,x}(t)$ represents the current supplied to the control coils which axis is paralel to the x axis of our system and $t_0$ represents the time step when the pulse is applied. In this case, the pulse has a duration of 0.005 s.

After that, when the sphere is at the maximum amplitude with respect to the equilibrium point (this will be the periapsis of the orbit), a second pulse is sent using the remaining pair of control coils (the pulse identical for each coil of the perpendicular pair). This second pulse places the sphere in the circular orbit. The current for the second pulse is identical to the first one,
\begin{equation}
	\label{eq:movcircular}
	I_{L,y}(t) = \left\{ \begin{array}{lcc}
		1.45\sin(180\pi(t-t_1)) & \mbox{if} & t_1 < t \leq t_1+0.005, \\
		\\ 0 & \mbox{otherwise},\\
	\end{array}
	\right.
\end{equation}
where in this case $t_1$ represent the time instant when the sphere is in its maximum amplitude in the x-axis movement.

\begin{figure}
	\centering
	\includegraphics[width=0.9\textwidth]{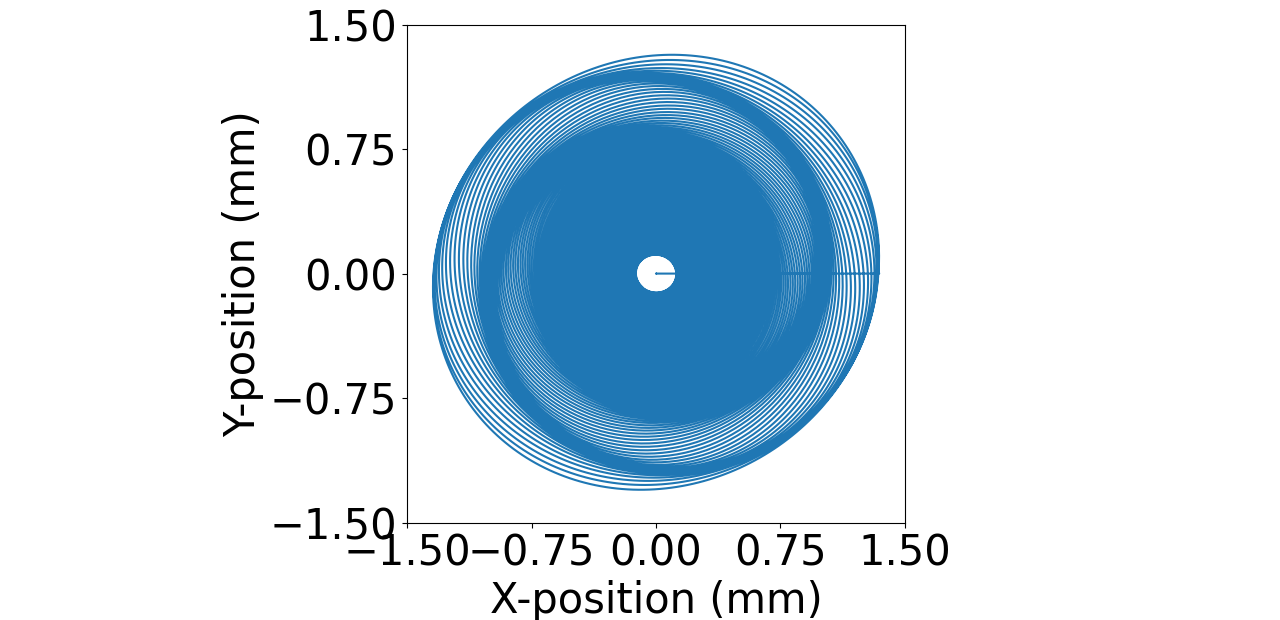}
	\caption{Circular movement}
	\label{circularmov}
\end{figure}

In our simulations we have solved the following Newton equation for the center of mass of the sphere,
\begin{equation}
  \frac{d^2{\bf r}}{dt^2}+\gamma\frac{d{\bf r}}{dt}=\frac{{\bf F}_{grav}+{\bf F}_{mag}}{m}
  \label{eq:motion}
\end{equation}
where $m=1.7\times10^{-6}$  kg is the mass of the sphere. Note that in the gravitational force we take into account the density of the fluid (128\,kg\,m$^{-3}$), and the damping parameter $\gamma$ is taken as $0.1$ s$^{-1}$ for He II at a temperature about 0.7\,K \cite{Schoepe95}. In Fig.~\ref{circularmov}, we plot the trajectory of this circular movement simulation for a time of 50\,s. 

As it may be noticed in the Fig.~\ref{circularmov}, the energy of the orbit decays with each cicle. This is because the damping parameter introduced in the dynamic equations has a great impact on this movement in the long term run. Hence, the speed of the sphere varies in each cycle, reaching its maximum value in the first cycles of the movement, which is approximately around $v_t=30$\,mm\,s$^{-1}$. Since we are interested in the motion of the sphere in stationary orbits at constant speed, it is necessary to compensate this energy loss introducing control over the motion of the sphere. One possiblility is to modify the movement introducing radial pulses when the trajectory of the sphere passes through each of the axis of the control coils.

\begin{figure}%
	\centering
	\includegraphics[width=0.8\textwidth]{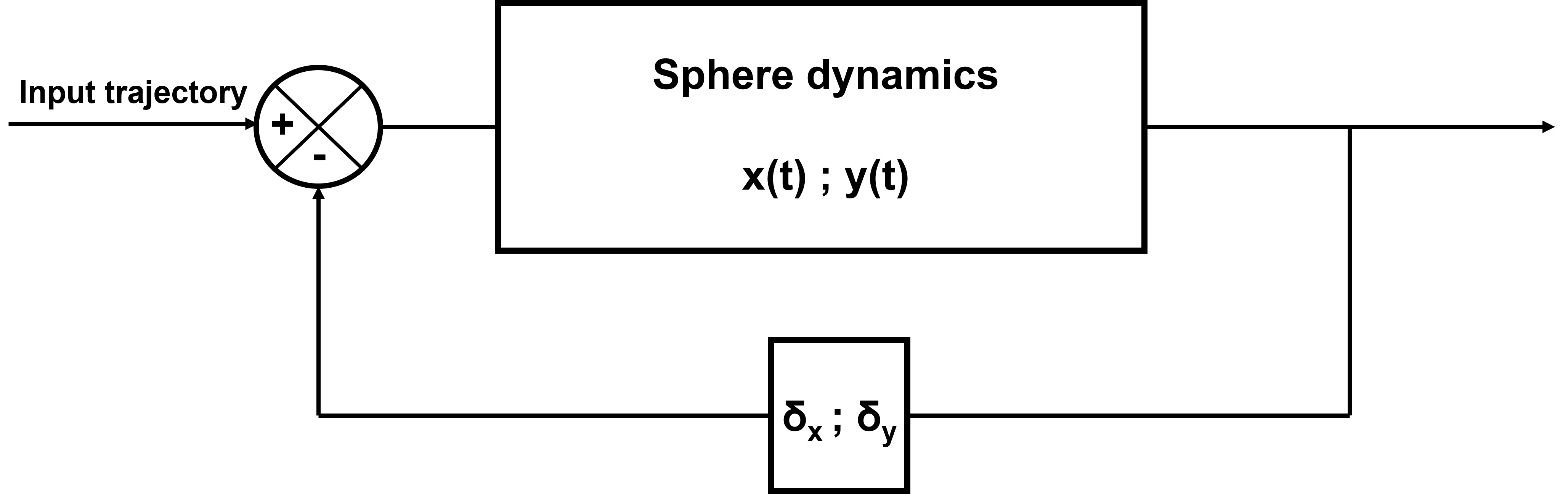}
	\caption{Negative feedback control system}
	\label{fig:negfeedback}
\end{figure}

To automate this control task we use negative feedback. The idea is to introduce in our system the desired trajectory (for simplicity let's suppose a circular trajectory, being the radius the only control parameter) and measure the error of the sphere's trajectory with the desired trajectory. Then, the control signal will depend on this error. In our system, we have considered a proportional control system, in which the control signal is directly proportional to the measured error $I_{C} \propto \epsilon$, where $\epsilon$ represent the difference between the desired position and the real position of the sphere. Hence, to control the sphere, we measure the position when the sphere passes through the axis of the control coils, i.e. when the sphere passes through the $x$ axis control coils, we compare the $x$ position of the sphere with the desired radius to evaluate the error in the $x$ axis, $\epsilon_x$  and the same is repeated for the $y$ axis, $\epsilon_y$. Once both errors are calculated, in the next period, a control pulse is applied to the sphere in the radial direction using the control coils at different instants,
\begin{equation}
	\label{eq:controlcircularx}
	I_{C,x}(t) = \left\{ \begin{array}{lcc}
		\delta_x\epsilon_x\sin(180\pi(t-t_{p,x})) & \mbox{if} & t_{p,x} < t \leq t_{p,x}+0.005, \\
		\\ 0 & \mbox{otherwise},\\
	\end{array}
	\right.
\end{equation}
\begin{equation}
	\label{eq:controlcirculary}
	I_{C,y}(t) = \left\{ \begin{array}{lcc}
		\delta_y\epsilon_y\sin(180\pi(t-t_{p,y})) & \mbox{if} & t_{p,y} < t \leq t_{p,y}+0.005, \\
		\\ 0 & \mbox{otherwise},\\
	\end{array}
	\right.
\end{equation}
where $\delta_x$ and $\delta_y$ are the proportional factors to control the $x$ and $y$ movement respectively. In addition, $t_{p,x}$ and $t_{p,y}$ represent the instants when the sphere passes through the $x$ and $y$ control coils axes. It is important to highlight that control pulses are only applied when the sphere crosses the x and y positive axes. The scheme of the negative feedback control system is depicted in \ref{fig:negfeedback}.

\begin{figure}
	\centering
	\includegraphics[width=0.8\textwidth]{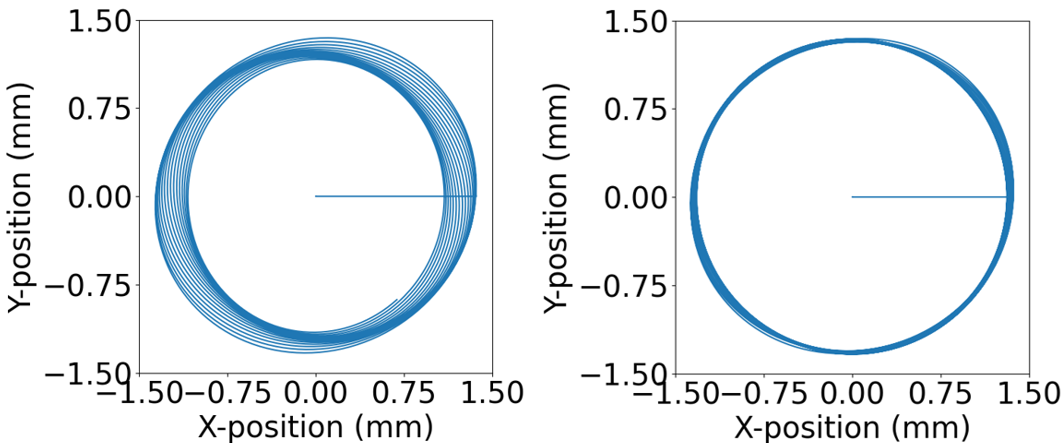}
	\caption{Comparison of t= 5\,s circular movement. The left-hand side graph represents the uncontrolled trajectory. In the right-hand side graph,  proportional control pulses are applied in every period in both pairs of the control coils as explained in the main text.}
	\label{fig:circcomparison}
\end{figure}

Finally, in Fig.~\ref{fig:circcomparison} we show a comparison between two simulations of the circular movement for 5\,s of computation. The left figure represents the movement without introducing the control scheme and the right picture represents the controlled movement.
In the controled set up, the control parameters $\delta_x$ and $\delta_y$ were both set to 300, and the desired radius was set to 1.4\,mm.

	\section{Preliminary experimental results} 
\begin{figure}
	\centering
	\includegraphics[width=0.65\textwidth]{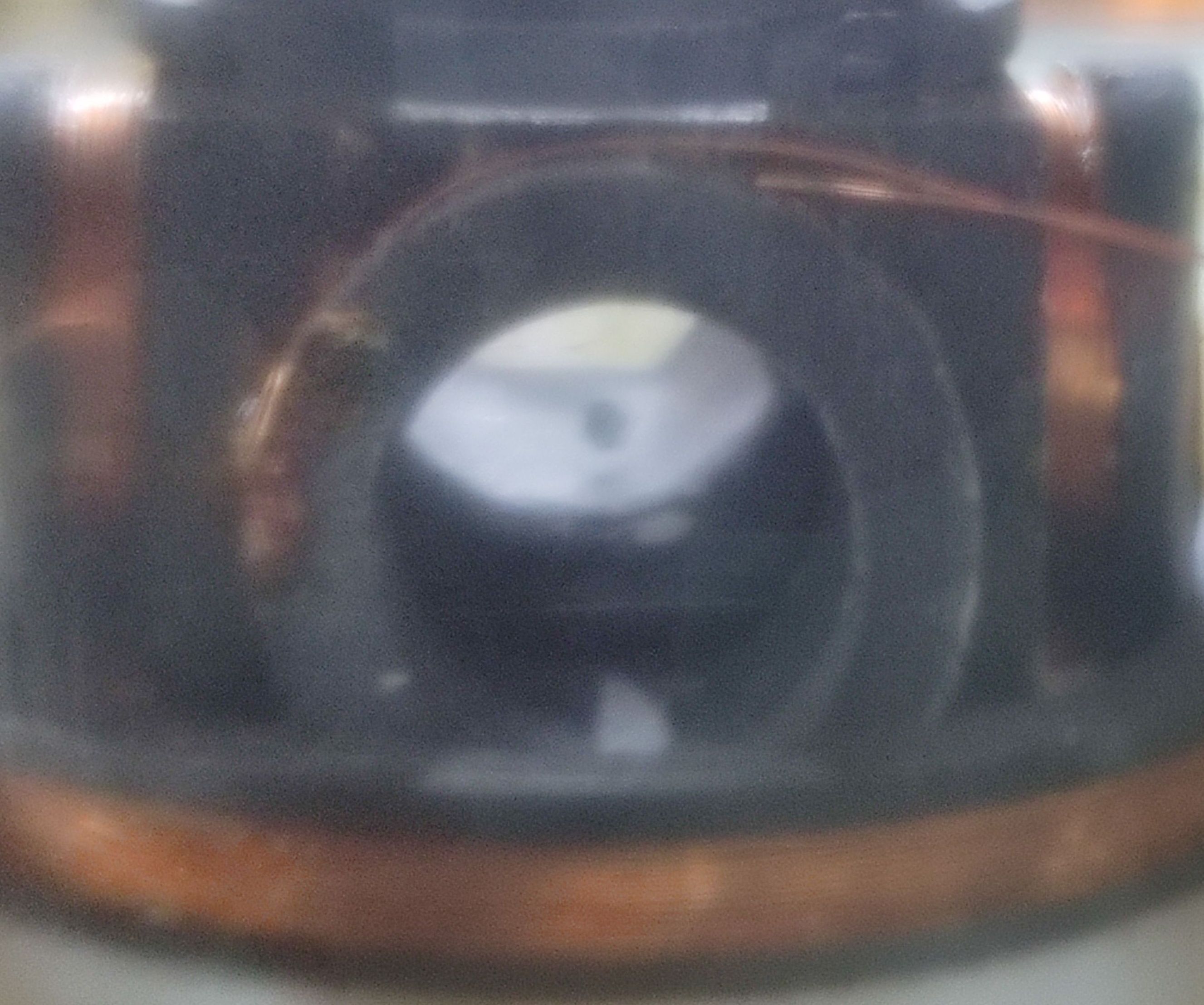}
	\caption{Indium-plated sphere levitating in superfluid $^4$He at $T=$1.5\,K}
	\label{fig:flying}
\end{figure}

In Fig.~\ref{fig:flying} we show the indium-plated sphere levitated in the middle of the experimental cell filled with superfluid $^4$He at $T=1.5$\,K. We have also managed to oscillate the levitated sphere by applying an AC current to the lateral coils along the $x$ axis \cite{Video}. The frequency was 2\,Hz and the driving current amplitude was 0.1\,A. It is clear that in this case the motion of the sphere is not harmonic. We intentionally used an elevated value for the driving current for demonstration purposes.

\section{Conclusions}
We have developed an apparatus for fine control of the motion of a superconducting sphere. The sphere can levitate in the bulk of a cryogenic fluid and is promising to be suitable for a wide range of measurements in both superfluid, normal and gaseous $^4$He and $^3$He \cite{Arrayas21}. Our finite elements analysis shows that the sphere can be driven in a variety of motion regimes, which will make a connection with numerous previous experiments in superfluids. Most importantly, the sphere can be made to move at a uniform velocity in a circle as well as in a straight line. This opens up a whole new multitude of approaches to quantitative studies of superfluid quantum matter, including quantum turbulence and dynamics of Andreev-bound states on the edges of topological superfluid $^3$He-B \cite{Autti20}, potentially hosting Majorana fermions. In these experiments the control over the quality of the surface will be crucial.

The sphere represents a very tractable geometry for numeric interpretation of the results of drag force measurement as a function of its velocity. However, in our design, the shape of the levitated object could be arbitrary, e.g. created using additive manufacture technology. This feature might  prove helpful for a facility utilizing cryogenic helium as a wind tunnel test fluid in classical turbulence experiments \cite{Donnelly01}. The advantage of our method is that the fluid remains stationary, removing the need to create a high-speed flow for the tests.

\section*{Dedication to Joe Vinen} The influence of Joe Vinen on the development of helium physics is difficult to overestimate. Our present work largely builds on the achievements made by him and his colleagues during many years of his active research. We had the pleasure of working with him on several projects, e.g.~\cite{Zmeev2015}, and retain very fond memories of Joe's commitment and insight in the field he genuinely loved.

\section*{Acknowledgments}
This work was funded by UK EPSRC (grant No. EP/P024203/1), the EU H2020 European Microkelvin Platform (Grant Agreement 824109) and by Universidad Rey Juan Carlos, Programa Propio (Analysis, modelling and simulations of singular structures in continuum models, M2604).

%


  
\end{document}